\documentstyle[preprint,revtex]{aps}
\begin{document}
\draft
\preprint{SNUTP-93\4}
\begin{title}
Non-Abelian Chern-Simons Quantum Mechanics
\end{title}
\author{Taejin Lee\cite{tlee}}
\begin{instit}
Department of Physics, Kangwon National University, Chuncheon 200-701,
KOREA
\end{instit}
\author{Phillial Oh\cite{poh}}
\begin{instit}
Department of Physics, Sung Kyun Kwan University, Suwon 440-746, KOREA
\end{instit}
\begin{abstract}
We propose a classical model for the non-Abelian Chern-Simons theory
coupled to
$N$ point-like sources and quantize the system using the BRST
technique. The resulting
quantum mechanics provides a unified framework for fractional spin,
braid statistics and
Knizhnik-Zamolodchikov equation.
\end{abstract}
\pacs{PACS numbers: 03.65.-w, 11.10.Ef}

\narrowtext

It is well known that charged particles in 2+1 dimensions acquire
fractional
spin and statistics through a long range interaction which can be
introduced
by the Abelian Chern-Simons term \cite{des}. These particles of fractional
spin and
statistics called anyons  play an important role in the theories for the
fractional quantum Hall effect and the high-$T_c$ superconductivity
\cite{wil}.

A non-Abelian generalization of anyons is also conceived; the particles
carry non-Abelian charges and interact with each other through the
non-Abelian Chern-Simons term \cite{des,witt,bal}. These particles,
which may be
called non-Abelian Chern-Simons (NACS) particles, acquire similarly
the fractional
spins and yet more generalized non-Abelian (or Braid) statistics \cite{fr}.
Recently it has been argued that the NACS particles also have
applications in the fractional quantum Hall effect \cite{moo} and their
wave functions can be interpreted as conformal blocks in certain rational
conformal field theories \cite{kz,ms}.

Although the classical action for the NACS particle is
constructed \cite{bal} and
some quantum mechanical aspects of them are discussed in the literature
\cite{witt,bal,ver,ems,bos,gua}, the precise connection between two
descriptions,
i.e., the classical and quantum, has never been presented. The purpose
of this letter
is to provide such a connection which may shed some light on the further
developement of the non-Abelian Chern-Simons quantum mechanics. To this end
we will thoroughly accomplish a canonical quantization of the classical
action for $SU(2)$ NACS particles and identify the gauge condition
which leads us
to quantum mechanical descriptions in the literature.

The classical Lagrangean for the $N$ NACS particles \cite{bal} can be
constructed in
terms of their spatial coordinates ${\bf q}_\alpha (t)\,\, ,\alpha = 1,
2,\dots$, $N$ and the isospin vectors $Q^a_\alpha (t)$ which transform under
the adjoint representation of the internal symmetry group. Defining
the isospin
vectors directly on the reduced phase space \cite{oh} which is $S^2$ for the
internal symmetry $SU(2)$,
\begin{equation}
Q^1_\alpha= J_\alpha \sin \theta_\alpha \cos\phi_\alpha,\quad Q^2_\alpha=
J_\alpha\sin \theta_\alpha \sin \phi_\alpha,\quad Q^3_\alpha =
J_\alpha\cos\theta_\alpha \label{iso}
\end{equation}
where $\theta_\alpha,\, \phi_\alpha$ are the coordinates
of the internal $S^2$ and $J_\alpha$ is a constant, one may write the
Lagrangean as
\[ L = \sum_\alpha\left(-{1 \over 2} m_\alpha \dot{\bf
q}_\alpha^2 +J_\alpha \cos \theta_\alpha \dot{\phi}_\alpha\right) -
\kappa\int
d^2 {\bf x} \,\epsilon^{\mu\nu\lambda} {\rm tr}\left(A_\mu \partial_\nu
A_\lambda +{2\over 3} A_\mu A_\nu A_\lambda\right) \]
\begin{equation}
+\int d^2{\bf x}\sum_\alpha \left(A^a_i(t,{\bf x}) \dot
q^i_\alpha +A^a_0(t, {\bf x})\right) Q^a_\alpha \delta ({\bf
x}-{\bf q}_\alpha)\label{lag}
\end{equation}
Here $\kappa=k/4\pi$, $k = {\rm integer}$, $A_\mu=A_\mu^a T^a$,
$[T^a, T^b] = \epsilon^{abc} T^c$, ${\rm tr} (T^a T^b) = -1/2 \delta^{ab}$,
and the space-time signature is $(+,-,-)$. The equations of motion from
the Lagrangean Eq.(\ref{lag}) contains Wong's equations\cite{wong}.

Since some of dynamical variables already appear as phase-variables
in the Lagrangean Eq.(\ref{lag}), we will not
introduce the canonical conjugates for them in the procedure of
Hamiltonian formulation
\cite{ber}.   Defining canonical momenta for the remaining variables
\[ p^i_\alpha ={\partial L \over \partial
\dot q_{i\alpha}}= m_\alpha \dot{q}^i_\alpha+ A^{a i} ({\bf q}_\alpha)
Q^a_\alpha\]
\begin{equation}
\pi^{a}_0 = {\partial L \over \partial \dot A^a_0}
= 0,\label{con}
\end{equation}
we find the first order Lagrangean
\[ L = \sum_\alpha\left(p^i_\alpha \dot q_{i\alpha} + J_\alpha
\cos \theta_\alpha \dot \phi_\alpha\right) +\int d^2{\bf x}\left(\pi^a
\dot A^a_0+  {\kappa\over 2}\epsilon^{ij}\dot A^a_i A^a_j\right) - H, \]
\[ H  = H_0 -\int d^2 {\bf x} \Biggl[  A^a_0\left( {\kappa \over 2}
\epsilon^{ij} F^a_{ij} +\sum_\alpha Q^a_\alpha \delta({\bf x}- {\bf
q}_\alpha)\right)+\lambda^a \pi^a\Biggr]  \]
\begin{equation}
 H_0 = \sum_\alpha {1\over 2 m_\alpha}\left(p^i_\alpha-A^{ai}
 ({\bf q}_\alpha) Q^a_\alpha\right)^2
\end{equation}
where we introduce  Lagrangean multiplier $\lambda^a$ to ensure the primary
constraint $\pi^a=0$.

The Euler-Lagrange equations can be reproduced as Hamiltonian equations
$\dot \xi = \{H, \xi\}$,  $\xi= q^i_\alpha, p^i_\alpha,  A^a_i,
A^a_0, \pi^a$,
provided the Poisson bracket is appropriately defined \cite{ber} as follows
\[ \{F, G\}= \sum_\alpha\left[\left({\partial F \over \partial
q^i_\alpha}{\partial G \over \partial p_{\alpha i}}-
{\partial F \over \partial
p_{\alpha i}}{\partial G \over \partial q^i_\alpha}\right) -{1\over
J_\alpha\sin\theta_\alpha}\left({\partial F \over \partial
\phi_\alpha}{\partial G \over \partial \theta_\alpha}- {\partial F \over
\partial \theta_\alpha}{\partial G \over \partial
\phi_\alpha}\right)\right] \]
\begin{equation}
 +\int d^2{\bf x}\left[\left({\delta F
\over \delta A^a_0({\bf x})} {\delta G \over \delta \pi^a ({\bf x})} -
{\delta F
\over \delta \pi^a({\bf x}) } {\delta G \over \delta
A^a_0 ({\bf x})}\right)+
{1\over\kappa}\epsilon_{ij}  {\delta F \over
\delta A^a_i ({\bf x})}{\delta G
\over \delta A^a_j ({\bf x})}\right]\label{poi}
\end{equation}
With the above Poisson bracket the fundamental commutators are given by
\[ \{q^i_\alpha, p_{\beta j}\} = \delta^i_j\delta_{\alpha\beta},\ \ \ \ \ \
\{Q^a_\alpha,Q^b_\beta\} =\epsilon^{abc}
Q^c\delta_{\alpha\beta}\]
\begin{equation}
\{A^a_i({\bf x}), A^b_j({\bf y})\} = {1\over \kappa}\epsilon_{ij}
\delta({\bf x}- {\bf y})\delta^{ab},\ \ \ \
\{A^a_0({\bf x}),\pi^b({\bf y})\} =\delta({\bf x}- {\bf y})\delta^{ab}.
\end{equation}

Viewing $\pi^a =0$ as a primary constraint, we obtain the Gauss's law as its
secondary constraint
\begin{equation}
\Phi^a = {\kappa \over 2}\epsilon^{ij} F^a_{ij} ({\bf x}) +
\sum_\alpha Q^a_\alpha \delta({\bf x}- {\bf q}_\alpha) =
0.\label{gauss}
\end{equation}
We also find that the constraints Eqs.(\ref{con})and (\ref{gauss}) form a
closed algebra
and no
further constraints arise
\begin{equation}
\{\Phi^a ({\bf x}),\Phi^b ({\bf y})\} =
\epsilon^{abc} \Phi^c\delta({\bf x}-{\bf y}), \ \ \ \ \ \ \
\{H_0, \Phi^a ({\bf x})\}=\{H_0, \pi^a ({\bf x})\} =
0.
\end{equation}

In order to discuss fixing the gauge degrees of freedom associated
with the first-class constraints Eqs.(\ref{con}) and (\ref{gauss}) and
a gauge independent
formulation,  will we resort to the (Becchi-Rouet-Stora-Tyutin) BRST method
\cite{brst}. Extending the phase space by introducing
two pairs of anti-commuting
ghosts $(b^a, c^a)$ and their canonical conjugates $(\bar c^a, \bar b^a)$,
\begin{equation}
\{\bar b^a ({\bf x}), c^b({\bf y})\} =
\{b^a ({\bf x}), \bar c^b({\bf y})\} = -\delta^{ab}\delta({\bf x}- {\bf
y}),
\end{equation}
we define the BRST charge $\Omega$ which is nilpotent
$\{\Omega,\Omega\}=0$ by
\begin{equation}
\Omega = \int d^2 {\bf x} \left( c^a\Phi^a -ib^a\pi^a +{1\over 2}
\epsilon^{abc} c^a c^b \bar b^c\right).
\end{equation}

The invariant action under the BRST transformation $ \delta \xi=
\{\xi,\Omega\},\,\,\,\xi= q^i_\alpha,\, p^i_\alpha,\,
Q^a_\alpha,\,  A^a_i,\,  A^a_0,\, \pi^a,\, b^a,\, c^a,\, \bar b^a,\,
\bar c^a $ is constructed to be
\begin{equation}
S=\int dt\left[K-H+\int d^2 {\bf x}\left(\pi^a\dot A^a_0+\dot c^a
{\bar b}^a+\dot b^a\bar c^a \right)\right]\label{action}
\end{equation}
where $H = H_0 -\{\Psi, \Omega\}$, $K=
\sum_\alpha\left(p_{\alpha i}\dot{q}^i_\alpha +J_\alpha \cos \theta_\alpha
\dot{\phi}_\alpha\right)+\kappa/2\int d^2 \,{\bf x}
\epsilon^{ij} \dot A^a_i
A^a_j$ and $\Psi$ is a Grassmann odd function of which explicit expression
depends on the gauge condition to be chosen.
The physical transition amplitude is represented by a path-integral
\begin{equation}
Z_\Psi =\int  D p D q D \cos\theta D \phi D A D \pi
D b D c D \bar b D \bar c \exp \{iS\}
\end{equation}
which is independent of $\Psi$.

The BRST formalism enables us to discuss the path integrals in various
gauges on an equal footing. However, we prefer to a gauge condition, if
exists, where the ghost variables are decoupled
from the physical degrees of
freedom and moreover the Gauss' constraints are solved explicitly. (In the
Coulomb gauge or in the covariant gauge the ghosts cannot be decoupled.)
There
is such a desirable gauge condition and it can be discussed best in the
complex coordinates where
\[ z = x+ iy,\quad \bar z = x- iy, \]
\begin{equation}
 A^a_z = {1\over 2} (A^a_1 - iA^a_2),\quad A^a_{\bar z} =
 {1\over 2} (A^a_1 +
iA^a_2),
\end{equation}
\[ z_\alpha = q^1_\alpha + i q^2_\alpha,\quad \bar z_\alpha =
q^1_\alpha - iq^2_\alpha.\]
The BRST invariant action Eq.(\ref{action}) is written in the
complex coordinates as
\[S =\int^{t_f}_{t_i} dt\left[K
 - H_0 +\{\Psi, \Omega\}+ \int d^2 z \left(\pi^a\dot
A^a_0+ \dot c^a{\bar b}^a+\dot b^a\bar c^a
\right)\right]\]
\[ H_0 = \sum_\alpha {2\over
m_\alpha}\left(p^{\bar z}_\alpha-A^{a}_z(z_\alpha, \bar z_\alpha)
Q^a_\alpha\right) \left(p^{z}_\alpha-A^{a}_{\bar z}
(z_\alpha, \bar z_\alpha) Q^a_\alpha\right). \]
\[ K = \sum_\alpha\left(p^{\bar z}_\alpha \dot z_{\alpha}+
p^{z}_\alpha \dot{{\bar z}}_{\alpha}+
J_\alpha \cos \theta_\alpha \dot \phi_\alpha\right) \]
\begin{equation}
+\int d^2 z \left(\pi^a \dot A^a_0+
{\kappa\over 2}\left(\dot A^a_z A^a_{\bar z} -\dot A^a_{\bar z} A^a_{
z}\right)\right)
\end{equation}

The path integral representing the physical amplitude is rewritten as
\[ Z =\int D p^z D p^{\bar z} D q^z D q^{\bar z} D\cos \theta D
\phi  D A_z D A_{\bar z} D \pi D A_0 \]
\begin{equation}
\exp \left\{-\kappa i \int d^2 z
\left(A^f_{\bar z} A^f_z + A^i_{\bar z} A^i_z\right)\right\}
\exp\{i\int^{t_f}_{t_i} dt
S\}
\end{equation}
where $A^f_z = A_z (t_f)$, $A^i_z = A_z (t_i)$. Note that we adopt the
coherent state quantization for the gauge fields.
Since $A_z$ and $A_{\bar z}$ are
treated as independent variables in the coherent state quantization,
we choose the gauge condition $A^a_{\bar z} = 0$ to fix the gauge degrees of
freedom. The corresponding gauge function
is given as
\begin{equation}
\Psi = \int d^2 z  \left({i\over \beta}\bar c^a A^a_{\bar z}
+\bar b^a A^a_0\right)
\end{equation}
where $\beta$ is a parameter which will be
taken to be zero at the end.
This gauge function is consistent with the BRST
quantization \cite{brst}. Since the gauge fields have only holomorphic
parts in this gauge, we may call it holomorphic gauge.

Scaling $\bar c^a \rightarrow \beta \bar c^a$, $\pi^a \rightarrow
\beta \pi^a$
and taking the limit $\beta \rightarrow 0$, we find
\[ S= \int dt\Biggl[ K -H_0 \]
\begin{equation}
+ \int d^2 z \left( i \bar c^a
D^{ab}_{\bar z} c^b + A^a_{\bar z} \pi^a -A^a_0\Phi^a -i \bar b^a b^a
-\epsilon^{abc} A^a_0 c^b \bar b^c\right)\Biggr]
\end{equation}
where $D^{ab}_{\bar z} = \partial_{\bar z}\delta^{ab}+ \epsilon^{abc}
A^c_{\bar z}$.
A further simplication can be made if the ghosts and $(\pi^a, A^a_0)$ are
integrated over. The integrations over the ghost variables $b^a$ and
$\bar b^a$ are Gaussian, so they can be trivially integrated out.
Integrations over $\pi^a$ and $A^a_0$ impose the gauge condition $A^a_{\bar
z} = 0$ and the Gauss' constraint $\Phi^a = \kappa F^a_{z \bar z}+
\sum_\alpha Q^a_\alpha\delta(z-z_\alpha) = 0$ respectively.
The remaining
ghost variables $c^a$ and $\bar c^a$ can be also integrated out, since they
produce in the path-integral ${\rm det} \partial_{\bar z}$ which is
independent of the dynamical variables. Thus the ghost are completely
decoupled from the physical degrees of freedom in the holomorphic gauge and
the resultant path integral is
\begin{equation}
Z  =\int D p^z D p^{\bar z} D
q^z D q^{\bar z} D\cos \theta D \phi
D A_z D A_{\bar z} \delta(A^a_{\bar z}) \delta
(\Phi^a) \exp\{i\int dt(K - H_0)\}
\end{equation}

The connection between the above path integral and the quantum mechanical
description becomes apparent if a few points are clarified further.
As advertized earlier, the Gauss' constraint can be solved explicitly
in the holomorphic gauge. Noting that the
Gauss' constraint in this gauge reduces to
\begin{equation}
-\kappa \partial_{\bar z} A^a_z + \sum_\alpha\left(
Q^a_\alpha\delta(z-z_\alpha)\right) =0,
\end{equation}
we easily obtain the solution
\begin{equation}
A^a_z (z, \bar z) = {i\over 2\pi \kappa}\sum_\alpha  Q^a_\alpha
{1\over z -z_\alpha} = {i\over 2\pi \kappa}\partial_z\left(\sum_\alpha
 Q^a_\alpha{\rm ln}(z -z_\alpha) \right).\label{sol}
\end{equation}
An operator version of the above solution  has
been presented in a discussion of the quantum holonomies acting
on the physical
state space of the Chern-Simons theory \cite{gua}.

With this solution we can discuss the quantum mechanics of NACS particles
without the gauge fields. Rewriting the path integral
representing the physical
amplitude
\[ Z = \int_{\xi_i}^{\xi_f} D[\xi, \pi_\xi] \exp
\{i\int^{t_f}_{t_i} dt (K- H)\} \]
\begin{equation}
 = <\xi_f\vert \exp\{ -i \hat{H}(t_f-t_i)\} \vert \xi_i>
\end{equation}
\[ \hat{H} = \sum_\alpha {2\over m_\alpha}
\hat p^z_\alpha \left(\hat p^{\bar z}_\alpha -  \hat A^a_z (
z_\alpha,\bar z_\alpha) \hat Q^a_\alpha\right)  \]
where $\xi = z_\alpha, \bar z_\alpha$, $\pi_\xi = p^z_\alpha, p^{\bar
z}_\alpha$ and the variables with ` $\hat{}$ ' denote quantum operators
$[ {\bar z}_\alpha, \hat p^{z}_\alpha] = i$, $[{z}_\alpha, \hat
p^{\bar z}_\alpha]= i$. Here the gauge field $ \hat A^a_z$ stands for the
operator version of the solution Eq.(\ref{sol}).
Since the iso-vector operators $\hat Q^a$'s satisfy the $SU(2)$
algebra, $[\hat Q^a_\alpha,\hat Q^b_\beta] =i\epsilon^{abc} \hat
Q^c_\alpha \delta_{\alpha\beta}$, they can
be represented by $SU(2)$ generators
$T^a_j$ in a representation of iso-spin $j$. For example, when $j=1/2$, we
represent $\hat Q^a$'s by the Pauli matrices $\tau^a/2$ and the state vector
$|\xi>$ by an iso-spin doublet. In passing, note that $\hat Q^2_\alpha=
J^2_\alpha = j_\alpha(j_\alpha+1)$, $j_\alpha\in {\bf Z}_{n+1/2}$.

Now we arrive at the quantum mechanical description of the NACS
particles whose
dynamics are governed by the Hamiltonian
\[ \hat {H} = -\sum_\alpha {1\over m_\alpha}\left(\nabla_{\bar
z_\alpha}\nabla_{z_\alpha}  +\nabla_{z_\alpha}\nabla_{\bar
z_\alpha}\right) \]
\[ \nabla_{z_\alpha} ={\partial\over \partial z_\alpha}  +{1\over 2\pi
\kappa}\left( \sum_{\beta\not=\alpha} \hat Q^a_\alpha
\hat Q^a_\beta {1\over
z_\alpha -z_\beta}+\hat Q^2_\alpha a_z (z_\alpha)\right)\]
\begin{equation}
\nabla_{\bar z_\alpha} ={\partial\over \partial \bar z_\alpha}\label{ham}
\end{equation}
where $a_z (z_\alpha)=\lim_{z\rightarrow z_\alpha} 1/(z-z_\alpha)$.

The second term and the third term
in $\nabla_{z_\alpha}$ are reponsible for the non-Abelian statistics
and the
anomalous spins of the NACS particles respectively. This can be seen
clearly, if
the wave function $\Psi_h$ for the NACS particles in the holomorphic
gauge is expressed
as follows
\[\Psi_h(z_1,\dots,z_N) = U^{-1}(z_1,\dots,z_N)
\exp\left(-{1\over 2\pi \kappa}\sum_\alpha
\lim_{z\rightarrow z_\alpha}\int^z
{\hat Q^2_\alpha \over z-z_\alpha} dz\right) \]
\begin{equation}
\Psi_a (z_1,\dots,z_N)
\end{equation}
where $U^{-1}(z_1,\dots,z_N)$ satisfies the
Knizhnik-Zamolodchikov equation \cite{kz}
\begin{equation}
\left({\partial\over \partial z_\alpha}  + {1\over 2\pi
\kappa} \sum_{\beta\not=\alpha} \hat Q^a_\alpha \hat Q^a_\beta {1\over
z_\alpha -z_\beta}\right) U^{-1}(z_1,\dots,z_N) =0.
\end{equation}
The KZ equation has a formal solution which is
expressed as a path ordered line integral in the
$N$-dimensional complex space
\begin{equation}
U^{-1}(z_1,\dots,z_N) = P \exp\left[-{1\over 2\pi\kappa} \int_\Gamma
\sum_\alpha d\zeta^\alpha  \sum_{\beta\not=\alpha}
\hat Q^a_\alpha \hat Q^a_\beta
{1\over \zeta_\alpha -\zeta_\beta}\right]
\end{equation}
where $\Gamma$ is a path in the
$N$-dimensional complex space with one end point fixed and the other being
$\zeta_f = (z_1,\dots,z_N)$. Explicit evaluation \cite{lee} of the
above formal
expression will give the monodromy matrices or the Braid matrices.

We see that $\Psi_a$ obeys
the non-Abelian statistics due to $U(z_1,\dots,z_N)$
while the wave function
$\Psi_h$ obeys the ordinary statistics and the Hamiltonian for
NACS particles becomes free Hamiltonian
in terms of $\Psi_a (z_1,\dots,z_N)$.
We also observe these particles carry fractional spin $2j_\alpha
(j_\alpha+1)/k$, because $\exp\left(-{1\over 2\pi \kappa}\lim_{z\rightarrow
z_\alpha}\int^z {\hat Q^2_\alpha \over z-z_\alpha} dz\right)$ acquires a
non-trivial phase $-{\hat Q^2_\alpha \over \kappa} i
= -2\pi i\left({2j_\alpha
(j_\alpha+1)\over k}\right)$ under the spatial $2\pi$ rotation.
In analogy with the Abelian Chern-Simons particle theory
we may call $\Psi_a$ the NACS particle wave function in the anyon gauge.
Therefore we have two equivalent descriptions for the NACS particles
as for the
Abelian Chern-Simons particles; in the holomorphic gauge and in the
anyon gauge. $U(z_1,\dots,z_N)$ is the (singular non-unitary) transformation
function between two gauges. It also defines the inner product in the
holomorphic gauge
\begin{equation}
<\Psi_1 |\Psi_2> = \int d^{2N}\zeta \Psi_1(\zeta)^\dagger U^\dagger
(\zeta) U(\zeta) \Psi_2 (\zeta)
\end{equation}
where $\zeta = (z_1,\dots,z_N)$. The Hamiltonian in the holomorphic gauge
Eq.(\ref{ham}) is certainly hermitian with this inner product.

It would be interesting to apply our work to
investigate thermodynamic properties of the NACS particles,
braid statistics and fractional quantum Hall effect.

\acknowledgements

T. Lee was supported in part by the KOSEF and in part by
non-directed research fund,
Korea Research Foundation (1992). P. Oh was supported by the KOSEF through
C.T.P. at S.N.U.


\begin{references}
\bibitem[*]{tlee}E-mail address:taejin@cc.kangwon.ac.kr
\bibitem[**]{poh}E-mail address:ploh@yurim.skku.ac.kr
\bibitem{des} S. Deser, R. Jackiw and S. Templetone,
Ann. Phys. (N.Y.) {\bf 140}, 372 (1982).
\bibitem{wil} F. Wilczek, ed., {\it Fractional statistics and anyon
superconductivity} (World Scientific, Singapore, 1990).
\bibitem{witt} E. Witten, Commun. Math. Phys. {\bf 121}, 351 (1989).
\bibitem{bal} A. P. Balachandran, M. Bourdeau and S. Jo,
Int. Jou. Mod. Phys. A {\bf 5}, 2423 (1990).
\bibitem{fr} J. Fr\"ohlich, in {\it Non-perturbative quantum field theory}
edited by G. 't Hooft {\it et al.} (New York: Plenum 1988).
\bibitem{moo} G. Moore and N. Read, Nucl. Phys. B {\bf 360}, 362
(1991); B. Blok and X. G. Wen, Nucl. Phys. B {\bf 374}, 615 (1992).
\bibitem{kz} V. G. Knizhnik and A. B. Zamolodchikov, Nucl.
Phys. B {\bf 247}, 83 (1984).
\bibitem{ms} G. Moore and N. Seiberg, Phys. Lett. {\bf 220B},
422 (1989); Commun. Math. Phys. {\bf 123}, 177 (1989).
\bibitem{ver} E. Verlinde, IAS preprint (1990)
IASSNS-HEP-90/60,  A note on Braid statistics and
the non-Abelian Aharonov-Bohm effect.
\bibitem{ems} S. Elitzur, G. Moore, A. Schwimmer and N. Seiberg, Nucl.
Phys. B {\bf 326}, 108 (1989).
\bibitem{bos} M. Bos and V. P. Nair, Int. J. Mod. Phys.
A {\bf 5}, 959 (1990).
\bibitem{gua} E. Guadagnini, M. Martellini, and M. Mintchev,
Nucl. Phys. B {\bf 336}, 581 (1990).
\bibitem{oh} P. Oh, Mod. Phys. Lett. A {\bf 7}, 1923 (1992).
\bibitem{wong} S. K. Wong, Nuovo Cimento {\bf 65A}, 689 (1970).
\bibitem{ber} F. A. Berezin, Comm. Math. Phys. {\bf 40}, 153 (1975);
J. R. Klauder, Phys. Rev. D {\bf 19}, 2349 (1979);
E. Witten, Comm. Math. Phys. {\bf 92}, 455 (1984);
L. Faddeev and R. Jackiw, Phys. Rev. Lett. {\bf 60}, 1692 (1988).
\bibitem{brst} For a review, see M. Henneaux, Phys. Rep. {\bf 126
}, 1(1985).
\bibitem{lee} T. Lee and P. Oh, in preparation.
\end{references}
\end{document}